\documentclass[10pt]{article}

\usepackage{amsmath}
\usepackage{amssymb}

\usepackage{graphicx}

\usepackage{cite}

\usepackage{color}

\usepackage[labelfont=bf,labelsep=period,justification=raggedright]{caption}

\makeatletter
\renewcommand{\@biblabel}[1]{\quad#1.}
\makeatother

\date{}

\begin{document}

\begin{flushleft}
{\Large
\textbf{Entropy involved in fidelity of DNA replication}
}
\\
\phantom{a}
J. Ricardo Arias-Gonzalez$^{1,2,3,\ast}$
\\
\phantom{a}
\bf{1} Instituto Madrile\~{n}o de Estudios Avanzados en Nanociencia (IMDEA
Nanociencia), Cantoblanco, 28049 Madrid, Spain 
\\
\bf{2} Department of Macromolecular Structure, Centro Nacional de
Biotecnolog\'ia (CNB-CSIC), Calle Darwin n$^{\circ}$3, 28049 Madrid, Spain 
\\
\bf{3} CNB-CSIC-IMDEA Nanociencia Associated Unit
``Unidad de Nanobiotecnolog\'{i}a" 
\\
$\ast$ E-mail: ricardo.arias@imdea.org
\end{flushleft}

\section*{Abstract}
Information has an entropic character which can be analyzed within the framework of the
Statistical Theory in molecular systems.
R. Landauer and C.H. Bennett showed that a logical copy can be carried out in the
limit of no dissipation if the computation is performed sufficiently
slowly. Structural and recent single-molecule assays have provided
dynamic details of polymerase machinery with insight into information processing.
Here, we introduce a rigorous characterization of Shannon Information in biomolecular
systems and apply it to DNA replication in the limit of no dissipation.
Specifically, we devise an equilibrium pathway in DNA replication to determine the
entropy generated in copying the information from a DNA template in the absence of
friction.
Both the initial state, the free nucleotides randomly distributed in certain
concentrations, and the final state, a polymerized strand, are mesoscopic
equilibrium states for the nucleotide distribution. We use empirical stacking free
energies to calculate the probabilities of incorporation of the nucleotides. The copied
strand is, to first order of approximation, a state of independent and non-indentically
distributed random variables for which the nucleotide that is incorporated by the
polymerase at each step is dictated by the template strand,
and to second order of approximation, a state of non-uniformly distributed random
variables with nearest-neighbor interactions for which the recognition of secondary
structure by the polymerase in the resultant double-stranded polymer determines the
entropy of the replicated strand.
Two incorporation mechanisms arise naturally and their biological meanings are
explained. 
It is known that replication occurs far from equilibrium and therefore the Shannon
entropy here derived represents an upper bound for replication to take place.
Likewise, this entropy sets a universal lower bound for the copying fidelity in
replication.


\newpage

\section*{Introduction\label{sec:intro}}

Many of the proteins in the cell are molecular motors which move along a molecular track and develop a mechanical work. Most of them work alone and therefore only an individual protein develops a
certain task without requiring or optimizing that task by working in cooperation. The single-molecule experimental approach to the study of these motors sheds light on their complex stochastic dynamics and its
connection to their biological function~\cite{Hormeno2006}.

Kinesin is probably the best characterized
molecular motor at the single molecule level~\cite{Svoboda1994}.
It is known that one of the roles of this protein is to transport cargoes along the
microtubules with high processivity, that is, to transport a cargo for long distances without detaching from the microtubular track. Polymerases on the other hand have a more complex task. They not only have to
translocate along a DNA template but most importantly, they have to copy a DNA single strand so that the fidelity in the so-called polymerization reaction is crucial for the cell division. To do so, DNA/RNA
polymerase actually works as both a Turing Machine and a Maxwell's
Demon~\cite{Maxwell1871,Szilard1929,Bennett1982,Berut2012}:
it is capable of successively reading one nucleotide at a time, identifying a complementary nucleotide
in the environment and
writing the information by catalyzing a phosphodiester bond in the nascent replicated strand. Moreover,
it is also capable of identifying errors in the copied strand by recognizing the secondary structure of the resulting double-stranded
polymer~\cite{Kamtekar2004,Julicher1998,johnson2004}.
Some of these proteins can correct a wrong nucleotide by removing
it and resuming the process in that position by the so-called proofreading
mechanism~\cite{Ibarra2009}, and some others include strand displacement
activity~\cite{Morin2012}.
DNA polymerase acts as a channel from the information point of view since it passes the genetic information
from a template strand to a copied one. The pairing process follows spontaneously by hydrogen bonding
and the emerging helical structure of the double-stranded polymer is mainly the result of the stacking
interactions between the new base-pair and its immediate previous neighbor in the polymer chain~\cite{SantaLucia2004}.

Kinesin uses the energy from the ATP hydrolysis to move along the microtubules in individual steps of 8 $nm$
by developing $5-8$ $pN$ forces, with an efficiency of $\sim 60\%$~\cite{Svoboda1994}.
More complexly, DNA polymerase uses part of the energy from deoxyribonucleotide triphosphates
(dATP, dCTP, dGTP and dTTP) hydrolysis for its own motion. Another part of the hydrolysis is used to branch
each incorporated nucleotide, that is, it is spent in the
phosphodiester bond formation which leads to the nucleobase incorporation in the nascent copied strand.
The remaining energy from the triphosphate nucleotides (dNTPs) hydrolysis plus that from the secondary
structure formation is still very high what makes paradoxically low the turnover efficiency of this enzyme ($\sim 23\%$)~\cite{Schliwa2003}. Besides, it is intriguing that the step of DNA polymerase is much shorter than that of kinesin (0.34 $nm$)
but the forces developed in each step are much higher ($\sim 10-30$ $pN$~\cite{Yin1995,Wuite2000}).

Although fidelity is the main role of this enzyme, the energy spent in accurately copying a single strand has only been included in the discussion of the energy balance
in the case of independent nucleotide incorporations~\cite{Wolkenshtein1961,Davis1965}.
Here we calculate the entropy that is needed to order free nucleotides in a reservoir by
following a sequence from a DNA template when no dissipation
is present. Our theoretical
framework allows the natural inclusion of interactions from near neighbors in the replication process. These interactions are closely related to the secondary structure formation of
double-stranded DNA and, subsequently, to error recognition by DNA polymerases~\cite{Kamtekar2004}.
On the light of this theoretical framework, we discuss the implication in the
energy comsumption by DNA polymerases.

DNA replication is a non-equilibrium process in which dynamical order is naturally
generated~\cite{Andrieux2008b}. Therefore, our calculation marks a lower
bound for the energy that must be spent in the ordering process otherwise
limiting polymerization.
As previously formulated~\cite{Andrieux2008b,Bennett1979},
a natural consequence of the present analysis is that DNA polymerase spends an energy in channeling information
from a template strand to a copied one with a fidelity which is increased in the presence of dissipation.
Our analysis allows envisioning how far from equilibrium this process occurs.

\section*{Theoretical framework\label{sec:theory}}
\begin{figure}[!ht]
\begin{center}
\includegraphics[draft=false,width=120mm]{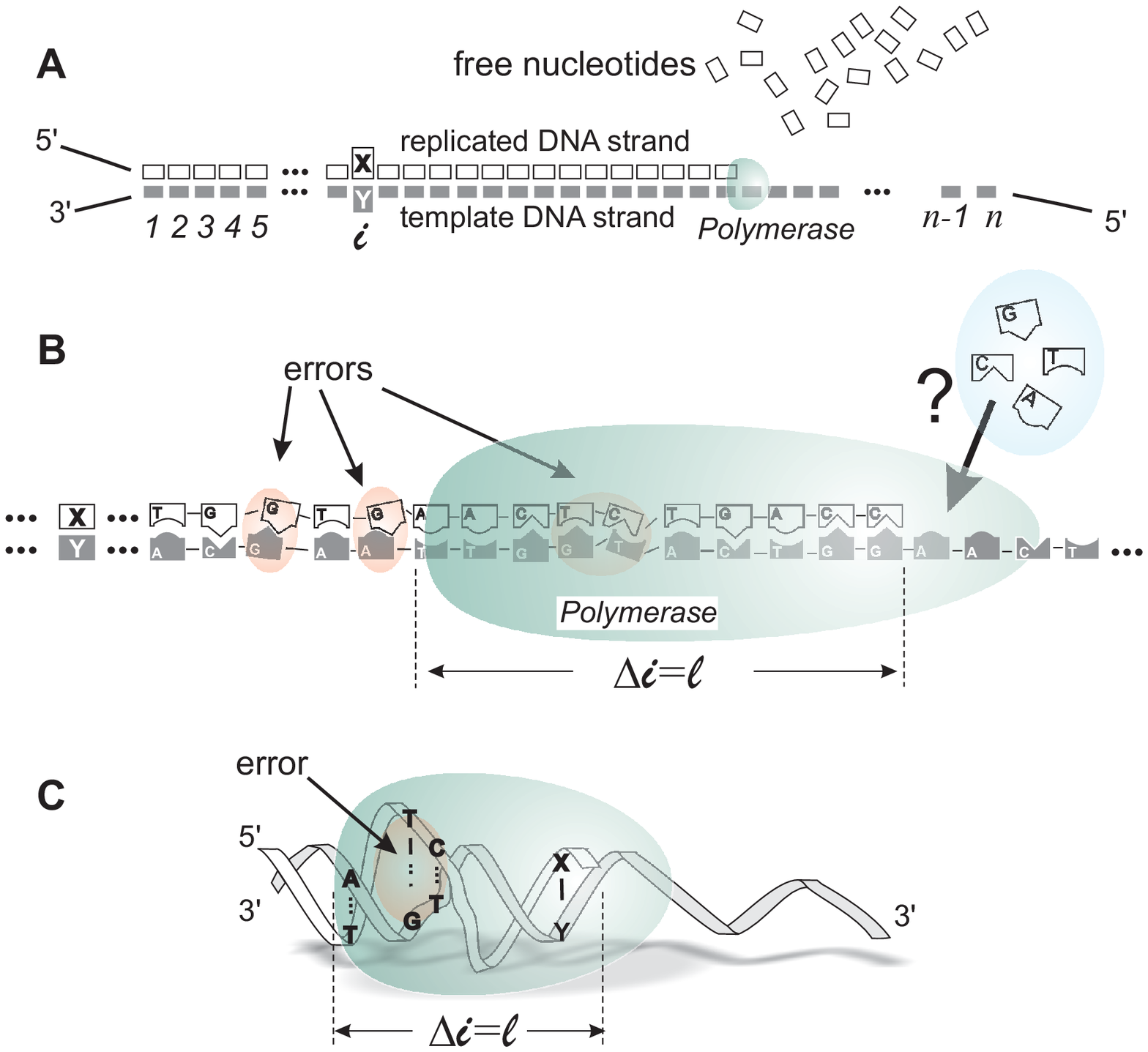}
\end{center}
\caption{{\bf Sketch of DNA replication.} (\emph{A}) Variables, $x$, parameters, $y$, and sequence position $i$ are represented together with 5' and 3' ends on both template and replicated strands. (\emph{B}) Polymerase (green) replicating a template DNA strand. Emphasis is placed on the linearity and directionality of the process.
$l$ is the number of nucleotides that this nanomachine covers when it is bound to the DNA.
(\emph{C}) Polymerase replicating DNA. Emphasis is placed on the recognition of errors based on the sequence-induced secondary structure of the DNA: the structure of DNA polymerase (a `palm',~\cite{Kamtekar2004}) and dsDNA
(a double helix) are evolutionarly adapted to optimally fit each other when Watson-Crick base-pairs
are formed. A complete turn of the double helix in B-form is $l\sim 10$ base-pairs.
\label{fig:SketchDNArep}}
\end{figure}

We start by developing a theoretical framework to analyze information transfer in
biomolecular systems.
As in former literature~\cite{Andrieux2008b,Andrieux2009,Klump2009}, we use a mesoscopic
approach to study genetic copying at the level of a single DNA polymer.
The process of ordering nucleotides according to a prescribed template sequence is shown schematically in
Fig.~\ref{fig:SketchDNArep}\emph{A}.
From a thermodynamic point of view, the initial and final states are mesoscopic
equilibrium states
although, as we study later in this article, the final state is different if the
copying process occurs in or out of
equilibrium~\cite{Ritort2008}.
We will state that a process occurs {\it in equilibrium} when it
takes place through an infinite number of small transitions between equilibrium states.
Then, we will say that the polymerase works {\it in equilibrium} when the nucleotide
selection and incorporation procedures performed by this enzyme takes place in the
absence of friction or other forces which irreversibly release
heat~\cite{Landauer1988,Bennett1973,Landauer1961}.
We suppose that the concentration of dNTPs is larger than that of pyrophosphate (PPi)
so that the process we study is phosphate-hydrolysis driven at all time but the motion
of the polymerase is
very slow so as to preserve equilibrium conditions.

A system which transitions between two equilibrium states
can increase its order if the system drives external energy through
appropriate dynamical paths~\cite{Gaspard2007}. In particular, Andrieux and
Gaspard~\cite{Andrieux2008b} showed that non-equilibrium temporal ordering
in copolymerization generates information at the cost of dissipation.
Based on experimental evidence from both structural and single-molecule studies,
here we mathematically model how DNA polymerase 'demon' channels energy from
dNTP hydrolysis to order nucleotides according to a template pattern by using
a minimum equilibrium description. Our scheme is an idealization of
real non-equilibrium copying mechanisms but will lead to a universal
(polymerase-independent) entropic upper bound for polymerization to take place.

A sequence in the template strand can be identified as a
vector of parameters $\textbf{y} = (y_1,\ldots,y_i,\ldots,y_n)$, where $i$ is an index which runs over the
nucleotide position and $n$ is the number of nucleotides in the template DNA strand. The copied
strand is represented here as a sequence of nucleotides given by the vector $\textbf{x} = (x_1,\ldots,x_i,\ldots,x_n)$ which stems from a multivaluate random variable $\textbf{X}$.
In replication, variables $X_i$ and parameters $Y_i$ take values over the same alphabet, namely ${\cal X} _{DNA} = {\cal Y} _{DNA} = \{A,C,G,T\}$. In transcription, they take values over isomorph alphabets, ${\cal X} _{DNA} = \{A,C,G,T\}$ and ${\cal Y} _{RNA} = \{A,C,G,U\}$. $A$, $C$, $G$, $T$ and $U$ stand for Adenine, Cytosine, Guanine, Thymine and Uracyl nucleotide class, respectively. Hence, we can express without loss
of generality for both replication and transcription: $x_i, y_i \in {\cal X}  = \{A,C,G,T\}$.

Since we are only dealing with the ordering process, we do not need to take into account the number of phosphates in the nucleotide or the oxidative state. In the initial state, the nucleotides are independent nucleobase entities with a triphosphate tail and in the final state, they are monophosphate molecular
subunits assembled in a linear chain by phosphodiester bonds.
However, the nucleobase information remains the same in both cases. In the case of replication, $A$, $C$, $G$ and $T$ are deoxynucleotides and in transcription $A$, $C$, $G$ and $U$ are oxynucleotides. Therefore, as mentioned, we neglect the chemical condition without loss of generality in the {\it informational} entropy analysis. Exact replication and transcription involve a
bijection between variables $X_i$ and $Y_i$ by the so-called Watson-Crick (WC) base-pairing rules,
but as we will see, non-WC unions can take place and give rise to copying errors.
In translation, the analysis is a bit more difficult since complementarity is replaced by the so-called genetic code which involves a surjective
correspondence between variables $X_i$ (individual aminoacids) and parameters $Y_i$ (triads of nucleotides).
This case will not be treated here.

The probability of having $n$ nucleotides in a certain sequence can be expressed as
$\text{Pr} \{X_1 = x_1, \ldots,  X_n = x_n \} = p(x_1, \ldots, x_n)$.
The corresponding entropy is, according to Gibbs formula,
$S(X_1, \ldots, X_n) = - k \sum_{x_1, \ldots, x_n} p(x_1, \ldots, x_n) \ln p(x_1, \ldots, x_n)$,
where $k$ is the Boltzmann constant, ``$\ln$" is the natural logarithm, and the random variables, $X_i$, take values, $x_i$, over the genetic alphabet $\cal X$. The calculation of these probabilities and their associated entropy involves the very complex analysis of the architecture of genomes and it is similar to that of generating text in a human language. Here, we calculate the entropy of copying the information from a
given DNA strand and therefore we are only dealing with the conditional entropy of the sequence
$\textbf{X}$ for a given sequence $\textbf{Y}$. Then, we simply need to use the probabilities of base-pairing nucleotides according to the template sequence, namely
$\text{Pr} \{X_1 = x_1, \ldots,  X_n = x_n || \textbf{Y} \} = p(x_1, \ldots, x_n|| \textbf{y})$,
where we have used a double bar to address the conditional character introduced by the complementarity in the base-pair formation. These probabilities can be expressed as a product of conditional probabilities in which each new, incorporated nucleotide, $x$, depends not only on the  template nucleotide, $y$, in front but also on
the previous base-pairs in the sequence,
$p(x_1, \ldots, x_n|| \textbf{y})  = \prod_{i=1}^{n} p\left(x_i | x_{i-1}, \ldots, x_1 || \textbf{y}_{(i)}\right)$, where $\textbf{y}_{(i)}$ is the parameter vector of the $i^{th}$ first nucleotides in the template strand. The entropy reads~\cite{Cover1991}:

\begin{eqnarray}\label{eq:entropy}
S(X_1, \ldots, X_n||\textbf{Y})
& = &
- k \sum_{x_1, \ldots, x_n} p(x_1, \ldots, x_n||\textbf{y}) \ln p(x_1, \ldots, x_n||\textbf{y})
\nonumber \\ [+3mm]
& = & 
- k \sum_{i=1}^{n}\sum_{x_1, \ldots, x_i} p\left(x_1, \ldots, x_i||\textbf{y}_{(i)}\right)
\ln p\left(x_i|x_{i-1}, \ldots, x_1||\textbf{y}_{(i)}\right)
\nonumber \\ [+3mm]
& = &
\sum_{i=1}^{n} S\left(X_i|X_{i-1}, \ldots, X_i||\textbf{Y}_{(i)}\right).
\end{eqnarray}

\noindent
The last part of this equation implies that the total entropy can be expressed as a sum over (double)
conditional entropies.

\subsection*{Polymerase supervision}\label{sec:PolSupervi}
\noindent
Polymerization is a spontaneous process at both room and physiological temperatures ($37^{\circ}C$)
since the free energies of nucleotide incorporation for WC base-pairing are negative. However, the process in the absence of
a catalyst may never occur. The biological catalyst or enzyme, the so-called polymerase, is not only able to accelerate the chemical reaction; it also has the capacity for recognizing the secondary structure in the
nascent double-helix polymer by a complex mechanism in which the polymerase structure is
involved~\cite{Kamtekar2004,Julicher1998,johnson2004}.
Its size covers approximately one helical turn of the double-stranded polymer and this determines a natural length or number of chained nucleotides, $l$, over which correct copying is supervised, as represented in Fig.~\ref{fig:SketchDNArep}, \emph{B} and \emph{C}.
One helical turn involves a number of nucleotides $l \sim 10$. This fact imposes a natural truncation over
the conditional probabilities of nearest base-pair neighbors. In other words, polymerase error recognition mechanism can be envisioned as a process in which this molecular machine supervises the copied strand by establishing correlations along $l$ previous base-pairs at each position, $i$, which mathematically involve
conditional probabilities. Then, the probability can be approximated by:

\begin{eqnarray}\label{eq:trunprob}
p(x_1, \ldots, x_n|| \textbf{y}) & \simeq &
p\left(x_1||y_1\right)p\left(x_2|x_1||y_2|y_1\right) \cdots
p\left(x_l|x_{l-1},\ldots, x_1||\textbf{y}_{(l)}^{(l)}\right) \times
\nonumber \\ [+3mm]
& &
\prod_{i=l+1}^{n} p\left(x_i | x_{i-1}, \ldots, x_{i-l} || \textbf{y}^{(l)}_{(i)}\right),
\end{eqnarray}

\noindent
where $\textbf{y}^{(l)}_{(i)}$ is the parameter vector of $l$ nucleotides which ranges between template positions $i-l$ and $i$. Eq.~\textbf{\ref{eq:entropy}} now establishes:

\begin{eqnarray}\label{eq:trunentropy}
S(X_1, \ldots, X_n||\textbf{Y})
& \simeq &
S(X_1|Y_1)+S(X_2|X_1||Y_2|Y_1)+ \cdots + S\left(X_l|X_{l-1}, \ldots, X_1||\textbf{Y}_{(l)}^{(l)}\right) +
\nonumber \\ [+3mm]
& &
\sum_{i=l+1}^{n} S\left(X_i|X_{i-1}, \ldots, X_{i-l}||\textbf{Y}^{(l)}_{(i)}\right).
\end{eqnarray}

Non-equilibrium paths which increase the fidelity of the copolymerization
process are ultimately determined by the above polymerase-DNA structural
fitting assumptions. Dynamical time evolutions are therefore concomitant to
the basic mechanisms which appear in equilibrium. Then, the
equilibrium description will provide an upper entropy bound for DNA
replication.

\subsection*{Entropy and Mutual Information}\label{sec:EntroMutInfo}
\noindent
The total entropy of the final state is
$S(\textbf{X},\textbf{Y})=S(\textbf{Y}) + S(\textbf{X}||\textbf{Y})$.
Parameters $\textbf{Y}$ have been fixed throughout evolution and therefore, we assume
within the polymerization problem that there is no uncertainty in determining these
parameters.
Hence, we set $S(\textbf{Y})=0$ without loss of generality. In these conditions, the
final entropy is given by the conditional entropy $S(\textbf{X}||\textbf{Y})$, as
expressed by Eqs.~\textbf{\ref{eq:entropy}} and~\textbf{\ref{eq:trunentropy}}.

The mutual information is
$I(\textbf{X};\textbf{Y})=S(\textbf{X})-S(\textbf{X}||\textbf{Y})$~\cite{Cover1991},
where $S(\textbf{X})$ is the entropy of the initial state. $S(\textbf{X})$ is the
entropy of the reservoir and it is
fixed for given nucleotide concentrations, as will be addressed later in this article.
Then, the lower the entropy of the final state, the higher the information acquired
in the copy, the higher the fidelity and the lower the number of copying errors.

Finally, the entropy change in the polymerization process, 
$\Delta S(\textbf{X}, \textbf{Y})=S(\textbf{X}||\textbf{Y})-S(\textbf{X})$,
and the mutual information are equal but opposite in sign in these conditions (note that
information is not defined in bits),
$I(\textbf{X};\textbf{Y})= -\Delta S(\textbf{X},\textbf{Y})$.

\section*{Results\label{sec:EntrFinalState}}
\noindent
The entropy in Eq.~\textbf{\ref{eq:trunentropy}} does not only depend on the number $l$ of nucleotides that are
imposed by the fitting length of the polymerase to the DNA template but also on the supervising
mechanism (e.g. see~\cite{johnson2004})
that the polymerase establishes by its architecture (molecular structure, e.g. see~\cite{Kamtekar2004}).
Due to the different polymerases that exist in nature and their diverse
structure and both polymerization and proofreading mechanisms, not mentioning
the cooperative associations of co-factor proteins in eukaryotic replication,
the calculation of the entropy cost of copying a nucleotide strand needs of a
heuristic model
to establish correlations over the $l$ nucleotides that the polymerase
supervises. Then, the energy spent in the ordering process involved in
polymerization is polymerase-dependent. However, a general upper bound
for the entropy
can be calculated for all the polymerases based on the fact that the secondary structure of the double-helix
of nucleic acids depends majorly on the immediate neighbors~\cite{SantaLucia2004}.
As a first approximation, we calculate the uppest bound by supposing that no influence of the previous
base-pairs exist ($l=0$). The picture of the polymerase within this approximation is that of a nanomachine which reads one nucleotide at a time and writes a complementary
nucleotide to the replicated strand. Later, we introduce the sequence-dependent
effects in an either, (1), reversible or, (2), irreversible copying process.

\subsection*{Entropy of nucleotides ordered with no neighbor influence\label{sec:ZeroOrder}}
\noindent
The picture of the polymerase within this approximation is shown in Fig.~\ref{fig:SketchDNArep}\emph{A}.
The polymerase only uses the information of one genetic symbol to decide the correct nucleotide to write in the replicated strand, and thus it is represented as if it only covered one position in the template strand.
In this case,
the random variables $X_i$ are independent although not identically distributed. Therefore, the probabilites and entropies are:

\begin{equation}\label{eq:prob0}
p(x_1, \ldots, x_n|| \textbf{y}) = \prod_{i=1}^{n} p(x_i || y_i),
\end{equation}

\begin{equation}\label{eq:S0}
S(X_1, \ldots, X_n|| \textbf{Y}) = \sum_{i=1}^{n} S(X_i || Y_i) =
\sum_{i=1}^{n} \sum_{x_i \in {\cal X}} p(x_i||y_i) \ln p(x_i||y_i).
\end{equation}

It is important to note that in general the probability, $p(x_i||y_i)$, depends implicitly on $i$ through the random variable $X_i=X(i)$ and explicity on the values of the parameters $y_i$.
The former dependence implies that
the copying process may be subjected to local property changes, such as nucleotide concentration or
temperature and ionic gradients, so that the polymerase position $i$ influences the probabilities.
The latter dependence
is explicit in the values of the parameters $y_i$ and addresses exclusively the sequence dependence along the template. We assume that polymerization is position-invariant
(cf. time-invariant random walk) and then we calculate the entropy from
Eq.~\textbf{\ref{eq:S0}} by using four independent probability distributions, namely
$p(x||y); x,y \in {\cal{X}} \Rightarrow \{ p(x||A), p(x||C), p(x||G), p(x||T) \}$.
Within this assumption, Eq.~\textbf{\ref{eq:S0}} becomes

\begin{equation}\label{eq:S0posInv}
S(X_1, \ldots, X_n|| \textbf{Y}) = \sum_{x,y \in {\cal X}} n_y p(x||y) \ln p(x||y),
\end{equation}

\noindent
where $n_y$ is the number of nucleotides of each type in the template sequence thus fulfilling
$\sum_{y} n_y =n$.
The Shannon entropy of a copied strand is no longer dependent on the template sequence but
on the number of nucleotides of each type on the template and its individual hybridization probabilities.
The transmitted
genetic information would be very poor if replication were taking place in the absence of nearest-neighbor base-pair interactions since a number of sequences $W(n) = \binom{n}{n_A,n_C,n_G,n_T}$ would be passing the
same information. If a dependence on position $i$ were explicit due to strong local property changes in
the environment, the transmitted information would be still poorer since the way boundary conditions affect
each replication reaction introduces a further uncertainty. The entropy of
nucleotides ordered with no neighbor influence was initially treated by Wolkenshtein and
Eliasevich~\cite{Wolkenshtein1961} and amended by Davis~\cite{Davis1965} to introduce wrong
incorporations. However these authors did not calculate the error probabilities which we
introduce next.  

The probability of incorporating a nucleotide $x$ in front of a nucleotide $y$ can be
estimated from experimental data as

\begin{eqnarray}
p(x||y) & = & \frac{1}{Z(y)}\exp{\left(\frac{- \Delta G^x_y}{k T}\right)},
\label{eq:prob0calc1} \\
[+3mm]
Z(y)    & = & \sum_{x \in {\cal X}}\exp{\left(\frac{- \Delta G^x_y}{k T}\right)},
\label{eq:prob0calc2}
\end{eqnarray}

\noindent
where $\Delta G^{x}_{y}$ is the energy released (negative) or absorbed (positive) upon pairing a nucleotide $x$ to another $y$ on the template strand and eventual stacking of the newly formed base-pair. Energies
$\Delta G^{x}_{y}$ are obtained from experimental data~\cite{SantaLucia2004} at $37^{\circ}C$
(polymerization occurs \textit{in vivo} in these conditions), as discussed in
\emph{Appendix A}. Then, the probabilities are:

\begin{equation}\label{eq:prob0matrix}
\textbf{P} = \left(p(x||y)\right) =
\left(\begin{array}{cccc}
0.054 & 0.012 & 0.035 & 0.765\\
0.033 & 0.011 & 0.789 & 0.034\\
0.144 & 0.961 & 0.118 & 0.128\\
0.769 & 0.016 & 0.058 & 0.072
\end{array}\right),
\\ [+3mm]
\end{equation}

\noindent
with $\sum_{x \in {\cal X}} p(x||y) = 1$. Note that the matrix elements follow the
order established by the alphabet sequence $x, y \in {\cal X}  = \{A,C,G,T\}$.
The WC base-pairs appear on the anti-diagonal and, as
expected, their propabilities are the largest. Although the real fidelity of a polymerase is much higher than
that represented by these probability values, many of its features are addressed by this matrix. Namely, as described in~\cite{Lee2006} for the human mitochondrial DNA polymerase, misincorporations which involve a $G$ are clearly favored and, with the exception of $G \cdot A$, the error $G \cdot T$ is the most common.
With the same exception, incorporations onto $T$ and $G$ are favored over $C$ and $A$.
As also measured by~\cite{Lee2006}, the error $C \cdot C$ is the least probable.
Moreover, it is found a discrimination between misincorporations $G \cdot T$ and $T \cdot G$.
A significant conclusion from this calculation is that errors in polymerization are mainly determined by the
thermodynamic affinity of nucleobases.

A similar discussion based on the raw free energy measurements instead of their associated probabilities
was formerly established in~\cite{Allawi1998}.
The role of kinetic and steric influence in replication fidelity and the importance of mismatch repair in error propagation was also therein discussed in the context of these thermodynamic data.
The effect of water exclusion in the active site of DNA polymerases has also been studied.
In particular, base-pair interactions were shown to be stronger than would otherwise be expected~\cite{Petruska1986, Petruska1988},
what should enhance the contrast among the probability values in matrix Eq.~\textbf{\ref{eq:prob0matrix}}.
Water exclusion in the DNA double helix has also been shown to decrease the axial base-stacking
interactions, as reflected in the
DNA stretch modulus~\cite{Hormeno2011}. It is therefore expected a type-dependent polymerase mechanism that
optimizes fidelity towards reported values of 1 error out of $\sim 10^5$
incorporated bases~\cite{Lee2006}.
The presence of exonuclease activity would enhance fidelity towards reported values of 1 error out of
$\sim 10^6 - 10^7$ incorporated bases~\cite{Lee2006}. Both reactions have been reported to be out of
equilbrium. Although not reaching these values, the entropy of the polymerized strand in equilibrium is
lower than that represented by Eq.~\textbf{\ref{eq:prob0matrix}} (i.e. the transmitted information is higher) due
to the presence of nearest-neighbor interactions between base-pairs. We analyze this influence in the next
section.

The entropy when no-neighbor interactions are present for a `class' of DNA templates
(i.e. with fixed $n_A$, $n_C$, $n_G$ and $n_T$) is calculated by
introducing the matrix values of Eq.~\textbf{\ref{eq:prob0matrix}} in Eq.~\textbf{\ref{eq:S0posInv}} and by using
the symbol parameters $y_i$, $i=1, \ldots, n$. A representative, template-independent value of the absolute
entropy of polymerization within this approximation can be calculated in the limit of uniform incorporation
of nucleotides. This calculation is performed in \emph{Appendix B} by using the formalism of
\textit{stationary random walk}~\cite{Cover1991} and the result is $s = 0.643$ $k$ per nucleotide ($k/nt$).

\subsection*{Entropy of nucleotides ordered within nearest-neighbor influence\label{sec:FirstOrder}}
\noindent
In this section we consider that the incorporation of a new nucleotide depends not only on the nucleotide
at position $i$ in the template but also on the recently formed base-pair at position $i-1$.
The physical nature of this dependence is the base-stacking interactions between base-pairs which make more probable to place a new nucleotide by a WC union than other combination since the eventual secondary structure of the resulting double-stranded polymer is more stable. The nearest-neighbor interactions implicitly
make both the probability and the entropy of the replicated strand become sequence-specific and increase the
fidelity of the transmitted information.
As pointed out before, the fact that a nearest-neihgbor approximation is sufficient to address secondary structure effects in the hybridization of two strands is supported by former
literature~\cite{SantaLucia2004}. Therefore we
introduce the hybridization energies from~\cite{SantaLucia2004} as coefficients $\Delta G^{y, y_{o}}_{x, x_{o}}$, which represent the free energy of positioning a nucleotide $x$ in front of a template nucleotide $y$ when the previously formed base-pair is $x_o$$-$$y_o$, to calculate the Shannon entropy of a replicated DNA
strand.

We assume that the polymerase is able to recognize the secondary structure of the double-stranded polymer,
as represented in Fig.~\ref{fig:SketchDNArep}\emph{C}, and thus decide the best match for each incorporated
nucleotide.
In doing this assumption, we use the fact that a polymerase is continuously grabbing nucleotides at random, fluctuating between an open and close conformation. The polymerase-dNTP binding energy stabilizes a close
conformation of the enzyme which is used to attempt the incorporation of each grabbed nucleotide to the template at each position $i$. Wrong matches are released in their initial triphosphate state and best
matches are hydrolyzed with release of PPi and eventually branched to the previously incorporated nucleotide
in the growing complementary strand through a polymerase-catalyzed phosphodiester bond.
The fluctuating state of the polymerase is restored after the nucleotide incorporation by using part of
the energy from the dNTP hydrolysis. This structural reset enables the enzyme to translocate to the
next template position~\cite{Yin1995,Guajardo1997} and leads to its memory
erasure~\cite{Bennett1973,Landauer1961,Berut2012}.

Polymerization can be considered a reversible reaction if the polymerase is not included in the process.
The inverse reaction, the logical unreading~\cite{Bennett1979} in which a branched nucleotide in its
monophosphate state (dMNP) is unbranched and released in a triphosphate state is the so-called
pyrophosphorolysis,
which is not biologically related to an editing process. The occurrence of this reaction depends on the concentration
of PPi in the reservoir, which we suppose to be low compared to that of dNTPs.
In the exonucleolysis reaction, which is performed by a different enzyme or by the
`exo' domain of some polymerases~\cite{Ibarra2009}, the initial state of the DNA template is recovered but
not that of the cleaved nucleotide since it is released in a monophosphate state with the consequent
dissipation of part of the energy from the phosphodiester bond breakage.
Exonucleolysis is thus irreversible, as expected from an editing process~\cite{Bennett1973,Landauer1961}.
However, having in mind that the reservoir is not affected by the substitution of a few dNTPs for dNMPs,
the total effect of polymerization and exonucleolysis can be assumed to encompass a reversible copying
process. This assumption will become clearer later.

Based on the previous-neighbor-influence assumption, the conditional probabilities are truncated for
$l=1$ in Eqs.~\textbf{\ref{eq:trunprob}} and~\textbf{\ref{eq:trunentropy}}. Then, the probability distributions are given
by coefficients $p(x_i|x_{i-1}||y_i|y_{i-1})$, such that
$\sum_{x_i \in {\cal X}}p(x_i|x_{i-1}||y_i|y_{i-1}) = 1$,
which implies that there is at least one nucleotide $x_i$ that binds to a nucleotide $y_i$ for a previously
formed base-pair made up of a nucleotide $x_{i-1}$ in front of $y_{i-1}$.
It is also assumed that the energies for nearest neighbors fulfill
$\Delta G^{x_{o}, x}_{y_{o}, y} = \Delta G^{y, y_{o}}_{x, x_{o}}$ (strand symmetry).
This assumption is true provided that hybridization energies do not show a higher-order dependence on the nearest neighbors within experimental
resolution~\cite{SantaLucia2004}. It is approximated otherwise~\cite{Huguet2010}.

The total energy of a configuration, $\nu$, made up of a sequence $\textbf{x}$ hybridized on a template
sequence $\textbf{y}$ is
$E_{\nu} \equiv E \left( x_1,\ldots, x_n || {\bf y} \right) =
\sum_{i=1}^{n} E\left( x_i|x_{i-1} || y_i|y_{i-1} \right)$
(see \emph{Appendix C}), where
$E\left( x_i|x_{i-1}||y_i|y_{i-1} \right)$ is the energy of pairing nucleotide $x_i$ on $y_i$ provided
that the previously incorporated nucleotide $x_{i-1}$ is already hybridized on the template nucleotide
$y_{i-1}$.
If the energies are not affected by local changes,
such as nucleotide concentration or temperature and ionic gradients,
their values will only depend on the position, $i$, on the template through the values of $y_i$.

The Shannon entropy can be calculated by using a partition function formalism, according to a hereafter labeled
as \textit{Ising mechanism}, or by using a Markov chain formalism, according to a hereafter labeled as
\textit{Turing mechanism}.

\subsubsection*{Ising and Turing mechanisms\label{sec:IsingTuring}}
\noindent
The Ising mechanism corresponds to an Ising model and it is thus calculated by using a partition function.
As we show below, the degree of reversibility of this mechanism depends on the stability of the base-pairs
as represented by their free energies.
The conditional probability at each step is given by (see \emph{Appendix C}):

\begin{equation}\label{eq:CondProbZ1}
p \left( x_i|x_{i-1}||y_i|y_{i-1} \right) = 
\frac{e^{-\beta E\left( x_i|x_{i-1}||y_i|y_{i-1} \right)}
      f\left( x_i||y_i, \ldots, y_n\right)}
     {\sum_{x'_i} e^{-\beta E\left( x'_i|x_{i-1}||y_i|y_{i-1} \right)}
      f\left( x'_i||y_i, \ldots, y_n \right)}
\end{equation}

\noindent
where

\begin{equation}\label{eq:f1}
f\left(x_i||y_i, \ldots, y_n \right) \equiv \sum_{x_{i+1},\ldots,x_n}
e^{-\beta E\left( x_{i+1}|x_i||y_{i+1}|y_i \right)}
\cdots
e^{-\beta E\left( x_n|x_{n-1}||y_n|y_{n-1} \right)}.
\end{equation}

The Turing mechanism, which as mentioned is based on the Markov chain formalism, is implicitly irreversible.
The probability of placing a nucleotide $x$ in front of a template nucleotide $y$ at position $i$ within this
formalism is given by (see \emph{Appendix C}):

\begin{equation}\label{eq:MarkovProb1}
p \left( x_i|x_{i-1}||y_i|y_{i-1} \right) =
\frac{e^{-\beta E\left( x_i|x_{i-1}||y_i|y_{i-1} \right) }}
{{\sum_{x'_i}} e^{-\beta E\left( x'_i|x_{i-1}||y_i|y_{i-1} \right)}}.
\end{equation}

Eq.~\textbf{\ref{eq:CondProbZ1}}
represents a probability which depends on the index $i$ through the symbol value $y_i$ and through the
length $n$ of the template strand chain.
Therefore, in the Ising mechanism, for which nucleotides are assumed to freely branch
and unbranch, the final macroscopic state is affected by the finite length of the genome.
In the Turing mechanism, Eq.~\textbf{\ref{eq:MarkovProb1}}, on the contrary, the probability only depends on the
position $i$ through the sequence. The latter is the process which takes place in polymerization in the
absence of exonucleolysis because it is associated to a unidireccional incorporation of nucleotides.
The former allows the already incorporated nucleotides to be replaced by new nucleotides and therefore it
naturally introduces the effect of exonucleolysis in equilibrium.
In the limit of very negative free energies for WC base-pairs
(high stability) and very positive free enegies for wrong base-pairs (low stability), the probabilities in
Eq.~\textbf{\ref{eq:CondProbZ1}} and~\textbf{\ref{eq:MarkovProb1}} become Kronecker delta-like functions
($p ( x_i|x_{i-1}||y_i|y_{i-1} ) = \delta^c_{x_i,y_i} \times \delta^c_{x_{i-1},y_{i-1}}$, where
$\delta^c_{x,y} = 1$ if $x$ is WC-complementary to $y$ and zero otherwise) and both
calculations converge to $S=0$, that is, information transmitted in the absence of errors.

\begin{figure}[!ht]
\begin{center}
\includegraphics[draft=false,width=70mm]{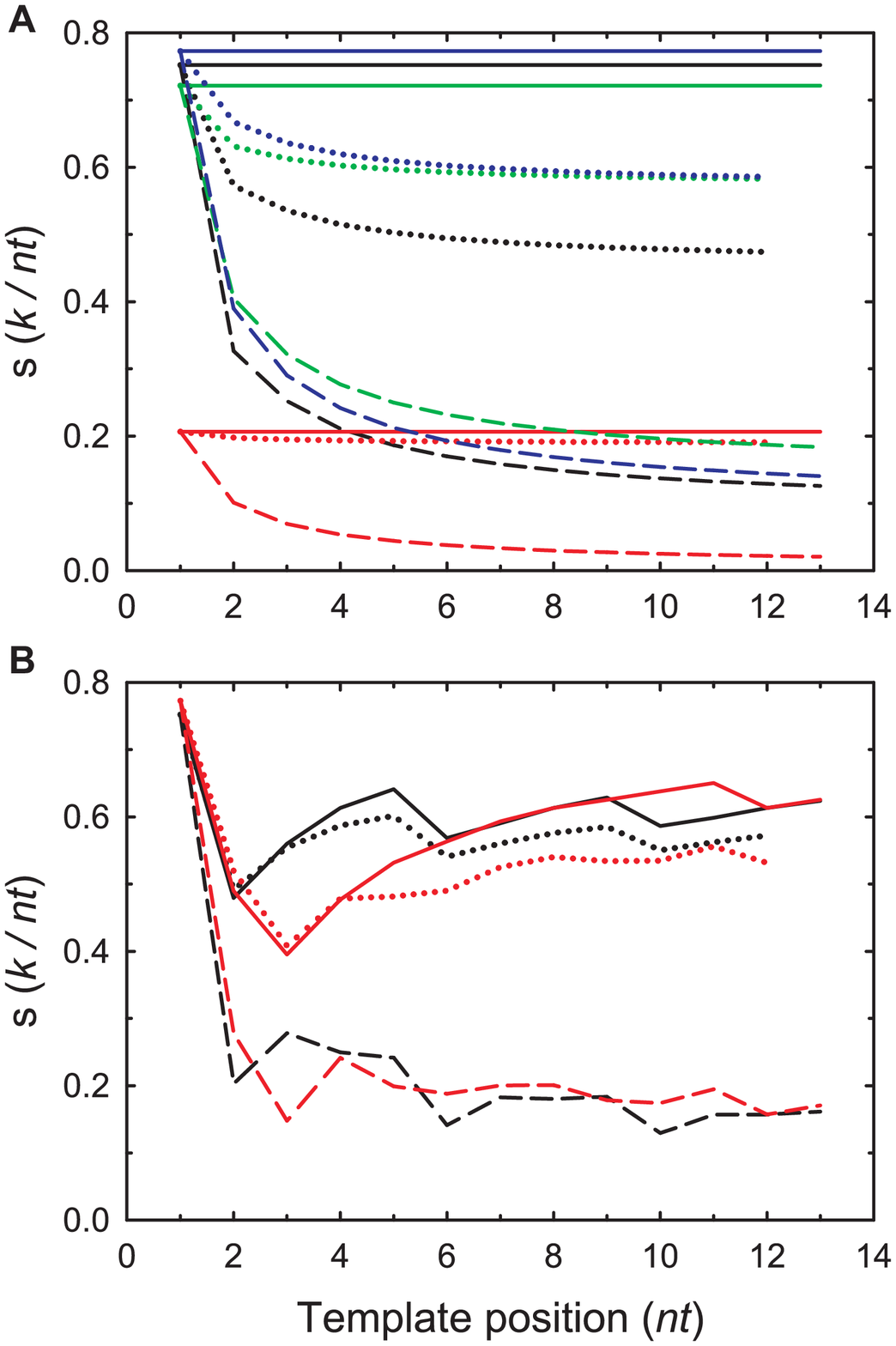}
\end{center}
\caption{{\bf Entropy of replication in equilibrium.} Each panel shows the entropy per nucleotide in the absence
(solid line) and presence of nearest neighbor influence for an Ising mechanism (dashed lines) and for a
Turing mechanism (dotted lines).
(\emph{A}) Monotonous template sequences: black lines, polydA; red lines, polydC; green lines, polydG; and blue lines,
polydT.
(\emph{B}) Black lines, periodic template sequence: ACGTACGTACGTA...; red lines, random template sequence
TCCGAGTAGATCT ...
\label{fig:Entropy}}
\end{figure}

\begin{figure}[!ht]
\begin{center}
\includegraphics[draft=false,width=70mm]{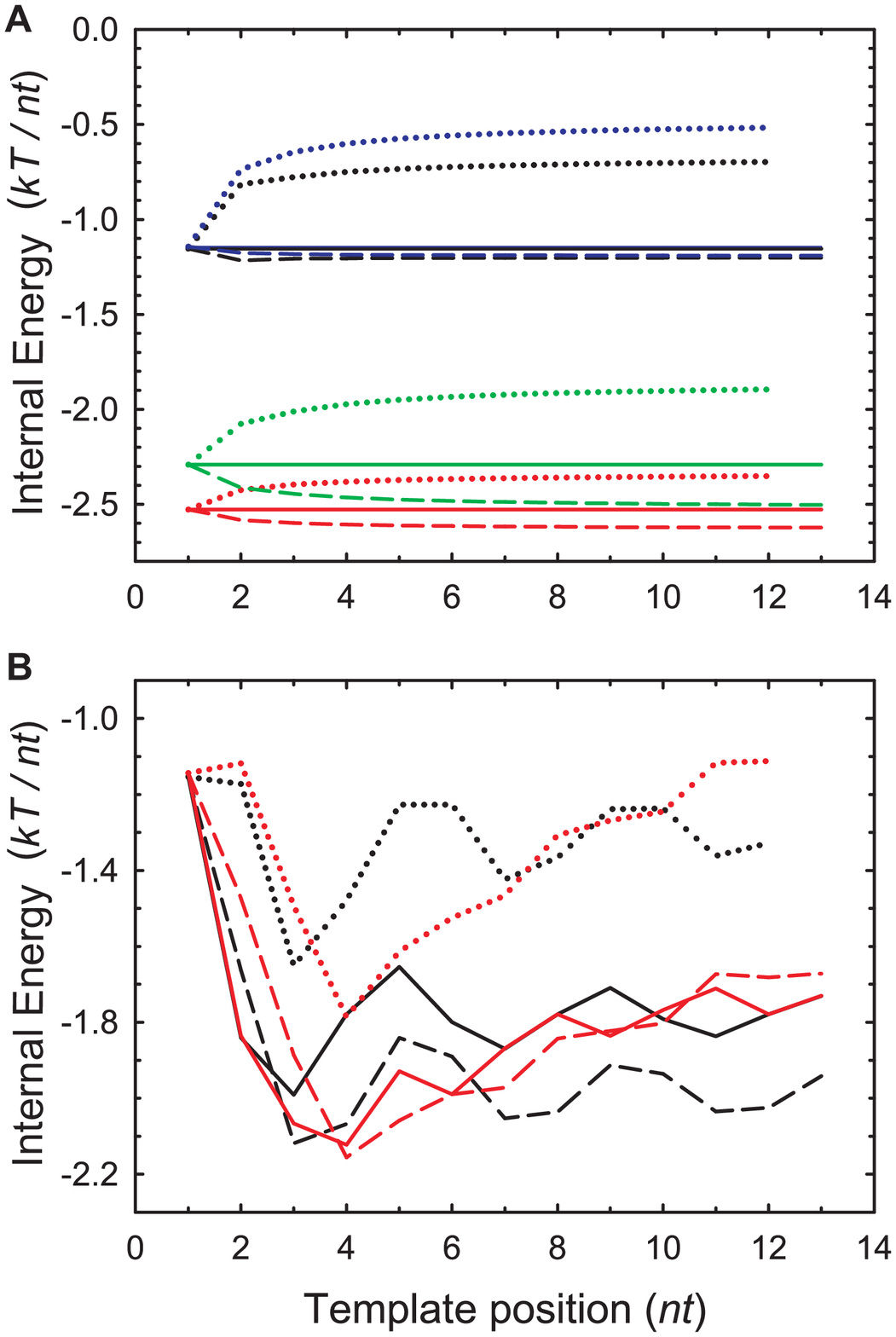}
\end{center}
\caption{{\bf Internal Energy of replication in equilibrium.} Each panel shows the mean energy per nucleotide in the
absence (solid line) and presence of nearest neighbor influence for an Ising mechanism (dashed lines) and for
a Turing mechanism (dotted lines).
(\emph{A}) Monotonous template sequences: black lines, polydA; red lines, polydC; green lines, polydG; and blue lines,
polydT.
(\emph{B}) Black lines, periodic template sequence: ACGTACGTACGTA...; red lines, random template sequence
TCCGAGTAGATCT ...
\label{fig:InternalEnergy}}
\end{figure}

\begin{figure}[!ht]
\begin{center}
\includegraphics[draft=false,width=70mm]{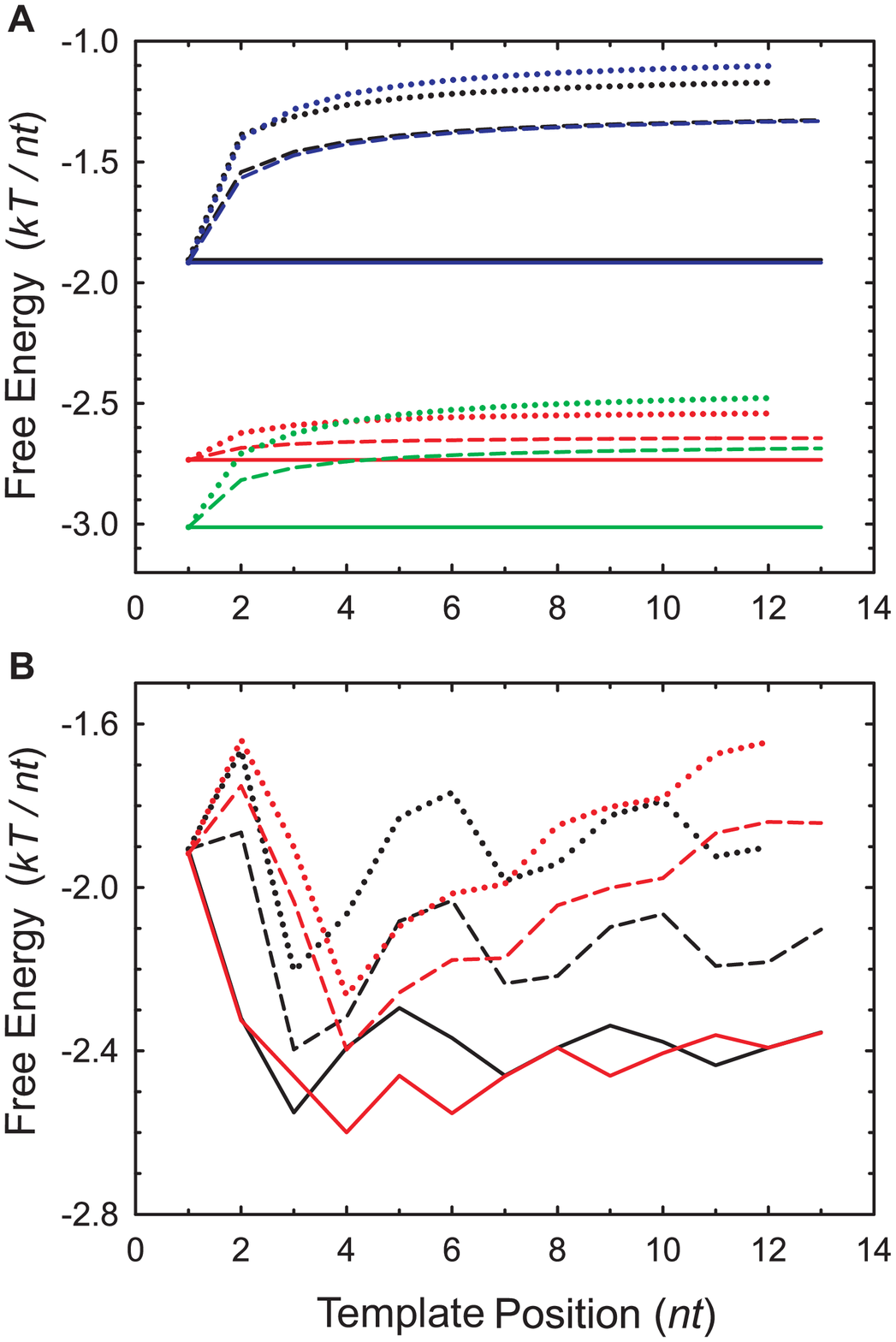}
\end{center}
\caption{{\bf Helmholtz Free Energy of replication in equilibrium.}
Each panel shows the free energy per nucleotide in the absence
(solid line) and presence of nearest neighbor influence for an Ising mechanism (dashed lines) and for a
Turing mechanism (dotted lines).
(\emph{A}) Monotonous template sequences: black lines, polydA; red lines, polydC; green lines, polydG; and blue lines,
polydT.
(\emph{B}) Black lines, periodic template sequence: ACGTACGTACGTA...; red lines, random template sequence
TCCGAGTAGATCT ...
\label{fig:FreeEnergy}}
\end{figure}

The entropy of replicating monotonous sequences, polydA, polydC, polydG, and polydT, and periodic and random
sequences is represented in Fig.~\ref{fig:Entropy} for both the Ising and Turing mechanisms.
As shown, the absolute entropy per incorporated nucleotide decreases and converges to the
thermodynamic limit very rapidly, within $\sim 13$ nucleotides, as further confirmed by Montecarlo
simulations of the internal energy (see \emph{Appendix D}).
For periodic sequences, the convergence reflects an attenuated periodicity which correlates with the template sequence. This trend is also observed for the internal energy, Fig.~\ref{fig:InternalEnergy}, and the Helmholtz free
energy, Fig.~\ref{fig:FreeEnergy}. The entropy when no neighbor interaction is
assumed is also plotted for comparison. This approximation for independent, non-indentically distributed
random variables is not dependent on the kind of calculation and the
resulting entropy is always higher than that obtained in the presence of nearest neighbor interactions.
This result is expected since when correlations are established among nearest neighbors which lead to
conditional probabilities, the probability of error decreases and the absolute entropy is closer to zero~\cite{Cover1991}.

Figure~\ref{fig:Entropy} also reflects another feature which takes place when
neighbor interactions are taken into account: the absolute entropy for the Ising mechanism is lower than for
the Turing one. Although in both mechanisms all configurations are accessible,
the number of pathways through which each configuration is accessible in the Turing mechanism is lower.
As stated above, the Ising mechanism is reversible and thus it naturally includes the effect of exonucleolysis,
what consistently leads to the lowest entropy and consequently to the lowest number of errors in
the replicated DNA, in agreement with previous proofreading analysis~\cite{Hopfield1974,Bennett1979}.
Finally, Fig.~\ref{fig:Entropy}\emph{A} reveals a large entropic
discrimination between polymerizing a polydG and a polydC which is not found between a polydA and a polydT.
This feature is consistent with what was shown for the case in which no neighbor interactions were taken into account (see matrix Eq.~\textbf{\ref{eq:prob0matrix}} in previous section).
This effect is purely entropic since the behavior of the internal and the Helmholtz free energies (Figs.~\ref{fig:InternalEnergy} and~\ref{fig:FreeEnergy}, respectively) do not exhibit such
discrimination.

\subsubsection*{Error rates\label{sec:Errors}}
\noindent
Errors are defined as non-WC unions. A gross estimation of the probability of error, $p_{error}$,
can be obtained through the Shannon-McMillan-Breiman theorem~\cite{Cover1991}, which states that
$-(1/n) k \ln p (x_1, \ldots, x_n||\textbf{y}) \rightarrow s (\cal{X})$ \textit{with probability 1}, where
$s (\cal{X})$ is the entropy per incorporated nucleotide (cf. entropy rate in a random walk).
Then, by defining the probability of error from the geometric mean
$1 - p_{error} \equiv p(x_1, \ldots, x_n||\textbf{y})^{1/n}$,
it follows that $p_{error} \rightarrow 1 - \exp (- s (\cal{X}) / \textit{k}  )$.
This definition implicitly assumes that each incorporated nucleotide is independent of the previous
base-pairs and therefore $p_{error}$ thus obtained represents a higher bound.
The entropy per incorporated nucleotide, as extracted from Fig.~\ref{fig:Entropy}\emph{A}
for general random sequences with equal
number of nucleotides of each class is $s \sim 0.1$ $k/nt$ for the reversible process, which leads to
$p_{error} \sim 10^{-1}$.

The probability of error can be more realistically calculated by simulating a large number of sequences according to the joint probability dictated by either the Ising or the Turing mechanism.
Montecarlo generation of sequences
(see \emph{Appendix D}) according to the templates
studied in Fig.~\ref{fig:Entropy} gives rise to $p_{error} \sim 10^{-2}$ for the Ising mechanism (the lowest average probability of error being for polydC template, $\sim 10^{-3}$) and $p_{error} \sim 10^{-1}$ for the
Turing mechanism (average probability of error for polydC, $\sim 10^{-2}$).
As a cross-check, we note that in the absence of nearest
neighbor interactions, these Montecarlo simulations give rise to an average
$p_{error} \sim 0.2$ (the lowest $p_{error}$, again, for polydC template, $p_{error} \sim 0.04$),
consistent with the information provided by matrix Eq.~\textbf{\ref{eq:prob0matrix}}.

The average probability of error therefore decreases, firstly, in the presence of nearest-neighbor interactions
and, secondly, for the Ising mechanism since, as explained above, this mechanism contains the effects of
exonucleolysis. Although the average probabilities of error in real, non-equilibrium replication, can be much
lower than the ones calculated here in equilibrium, similar error rates have been reported for some
polymerases~\cite{Loeb1982}.

\subsection*{Internal and Free Energies\label{sec:UandF}}
\noindent
Figure~\ref{fig:InternalEnergy} shows the behavior of the mean energy (see \emph{Appendix C})
which is released upon incorporating a new nucleotide at each step of the polymerase.
The stored information in the double-stranded polymer gives rise to a higher
internal energy in absolute value for the Ising mechanism than for the Turing one since the number
of WC unions, which involve stronger interactions than other pairing possibilities,
is higher for the former mechanism.

An error decreases the stability of a microstate (i.e. decreases in absolute value the (negative) energy,
$E_{\nu}$, of an individual nucleotide arrangement, $\nu$, in the copied strand) with respect to the case of
a correct (WC) match, not only by contributing with an either less negative or positive energy at the position of
incorporation, $i$, but also at the next step, $i+1$, independently of whether the next incorporated
nucleotide is a correct match or another error. This does not happen for the case in which no nearest-neighbor
influence is taken into account since in that case an error only affects the stability of a microstate
at the position where the wrong nucleotide is incorporated.
Therefore, if the number of errors when the influence of previous
neighbors is taken into account is not much smaller than for the case in which no influence is taken
into account, the total energy of a microstate will be on average lower
(i.e. will give rise to a less stable configuration) for the former.
This is why the internal energy for the Turing mechanism is lower in absolute value compared to
that in the absence of nearest-neighbor interactions.
The internal energy for the Ising mechanism is however higher in absolute value than for the case in which
no nearest-neighbor interactions are taken into account because in this mechanism the number of errors is
much smaller than for the Turing one (see Fig.~\ref{fig:InternalEnergy}).

The Helmholtz free energy, Fig.~\ref{fig:FreeEnergy}, reflects that the information generated under
the Ising mechanism is more significant than that generated under the Turing one since the former produces a
smaller number of errors. The free energy also provides information about the spontaneity of copying
a template DNA strand. As shown, for a given initial free energy it is more favorable to write a DNA replicate
under a process in which each copied symbol does not have a memory of the copying history.
When neighboring interactions are considered, the process for which nucleotides are not written by following
a directionality (Ising mechanism) is favored over that in which the symbols must be copied on a directional
one-after-one basis (Turing mechanism) in the 3' to 5' template sense.
It is important to note however that the real, non-equilibrium process involves a
more complex enzymatic coordination for the former procedure, what actually involves a different physical
pathway that could make such procedure become more unfavorable
(e.g. see~\cite{Loeb1982,Kamtekar2004,Ibarra2009}).

\subsection*{Initial entropy of the nucleotides from the reservoir\label{sec:EntrInitState}}
\noindent
The entropy of the initial state is determined by a distribution of probability that
a certain nucleotide is grabbed by the polymerase. Hence, this entropy addresses the
order in which individual nucleotides reach the `pol' site of the enzyme with
independence of whether they would be eventually incorporated to the replicated
strand or discarded back to the reservoir.
This entropy is not unique and depends majorly on the concentrations of the nucleotides. For a reservoir saturated with each nucleotide class, the entropy can be calculated by the Boltzmann formula,
$S_0 = k \ln W(n)$, where $n$ is the number of microstates compatible with a `macrostate' $W$. The nucleotide numbers, $n_x$, $x \in {\cal X}$, fulfills $\sum_{x} n_x =n$ and therefore $W(n)$ is given by the multinomial coefficient. As expected, for large $n$, Gibbs and Boltzmann formula provide the same values, namely

\begin{equation}\label{eq:asymptotic}
\ln W(n) = \ln \frac{n!}{n_A! n_C! n_G! n_T!} \sim -n \sum_{x \in {\cal X}} p_n(x) \ln p_n (x)
\end{equation}

\noindent
where $p_n(x) = \frac{n_x}{n}$ is the probability of grabbing a nucleotide of type $x$ which is in the reservoir at concentration $n_x/n$. A similar approach to this entropy can be extracted from the case of the
ideal gas. The {\it informational} contribution to the entropy for this system is the same, as expected,
and has been analyzed in \emph{Appendix E}.

By setting equal concentrations for all the nucleotide types, the initial entropy is $s_0 = k \ln 4 = 1.39$
$k / nt$. This is the value addressed by setting $W(n)=4^n$, which better describes the infinite
number nucleotides of each class contained in an ideal reservoir. The entropy difference between the initial
and final states is therefore $\Delta s \simeq -1$ $k / nt$. This entropy change
cannot be much larger
than this value since the lowest (non-equilibrium) final entropy is in any case $S>0$. However, approaching
$S=0$, or zero error rate, involves an ever increasing energetic cost with subsequent
dissipation~\cite{Andrieux2008b,Bennett1979}, in accord with the third law of thermodynamics.

\section*{Discussion\label{sec:discussion}}
\noindent
A common mechanical action of linear molecular motors such as kinesin and polymerases is translocation along
a molecular track. However, the main role of polymerases is the copying fidelity of the DNA,
being this double-stranded polymer
a particular molecular track which stores information. The selection of one correct
nucleotide at each translocating step of the polymerase constitutes a mechanism which needs of energy as well.
We have calculated the entropy balance of a system of nucleotides randomly distributed in a reservoir which are
finally incorporated into a copied strand according to a template DNA in the absence of energy dissipation.
To that end, we have evaluated the Shannon entropy at both the initial and the final states of the nucleotide
symbols by connecting these states through an equilibrium pathway.
We show that the entropy related to fidelity must be reduced from the initial state in $\simeq 1$ $k$ at
each step of the polymerase.
Given that the initial internal energy of the free nucleotides is $(3/2)$ $kT / nt$ (equipartition theorem),
their associated entropy is $\simeq 1.4$ $k / nt$, and that the final free energy is that shown in Fig.~\ref{fig:FreeEnergy}, the free energy invested in copying fidelity must be at least $\simeq 2$ $kT / nt$.

A gross analysis of the bulk chemical equilibrium between correct/incorrect incorporation of
nucleotides shows that the energy required to maintain a copying fidelity of one wrong nucleotide out
of $10^m$, i.e. an error rate $p_{error} \sim 10^{-m}$, is
$\Delta G = - k T \ln p_{error} = 2.3m$ $k T$~\cite{Loeb1982},
similar to but larger than the above analysis for a low number of errors (i.e. for $m=1, 2$ and $3$)
since in this estimation the polymerase is not assumed to work very slowly.
In particular, for real error rates of $\sim 10^{-3} - 10^{-7}$, the free energy is much larger than 2 $kT/nt$.
If this energy is added to the thermodynamic efficiency~\cite{Parrondo2002} of
polymerases, the resulting values would be much higher than those estimated from just the
analysis of the translocation mechanism~\cite{Schliwa2003}.
The contrast between this analysis and the equilibrium polymerization scheme presented in this article
demonstrates, on the one hand, that the copying pathway in polymerization (which may be coupled to the translocation one) is
far from equilibrium, and on the other hand, that the final state of the nucleotides in the copied strand depends on whether the copying mechanism ocurrs in or out of equilibrium. The latter implies that information managing results in very different fidelities depending on how far from equilibrium the copying mechanism takes place, in agreement with former literature~\cite{Andrieux2008b}, and on which the specific polymerase
mechanism is. 

Non-equilibrium paths can certainly destroy the information
acquired in equilibrium, but they can also amplify it. We have explained that
DNA-polymerase structural fitting is responsible for increasing dynamical
order in the replication process.
Then, the analysis presented here shows a universal higher bound of absolute
entropy in polymerization and,
subsequently, an error tolerance for the copying fidelity.
Each individual polymerase actually uses a specific replication mechanism in the presence or absence of
exonucleolysis which sustains an associated error rate evolutionarily coupled to its cellular line development. In our analysis, we include the effects of the previous neighbor base-pair whose physical nature
is the base-stacking interactions. These interactions are responsible for the secondary
structure of DNA ---the double helix--- and its correct formation is supervised by the DNA polymerase
through structural fitting~\cite{Kamtekar2004,Julicher1998,johnson2004}.
We show that such supervising mechanism reduces the entropy of the copied strand
with respect to the case in which these interactions are neglected, a consequence of the fact that
information fidelity increases in the presence of conditional
probabilities~\cite{Cover1991}.

Finally, we show that the inclusion of the nearest neighbor interactions
leads to different absolute entropies of the polymerized strand depending on
whether nucleotides are incorporated in either an irreversible or a reversible
process. The latter presents the lowest absolute
entropy, which is consistent with the error reduction generated by
proofreading~\cite{Hopfield1974,Bennett1979}, a mechanism in which
nucleotides are removed by exonucleolysis in a backtracking motion of the
polymerase or by the presence of
an exonuclease enzyme. Error rates within these two equilibrium mechanisms
with nearest-neighbor influence are in the $10^{-1}-10^{-3}$, better than the
most simple scheme where these interactions are neglected and near some real
polymerase fidelities~\cite{Loeb1982}.
Most commonly reported polymerases however strongly differ from these rates
what ultimately reflects how far from equilibrium they work.
The equilibrium mechanisms described here are inherent to more general
non-equilibrium polymerization pathways since time
evolutions are ultimately mediated by the polymerase demon action.

Although polymerases speed up the replication/transcription reactions, it is
important to note that translocation for these molecular motors is slower than
for transport molecular motors such as kinesin and myosin~\cite{Schliwa2003}.
This fact suggests that non-equilibrium replication pathways are mainly focused
on the regulation of specific error rates in copying fidelity rather than in
the translocation mechanism.

\section*{Conclusions\label{sec:conclusions}}
\noindent
We have conceived a probabilistic framework based on structural and
single-molecule experimental results which models
the copy of genetic information by molecular motors through the recognition of
DNA secondary structure. The link between thermodynamic entropy, which is based on statistical concepts at the molecular level, and Shannon entropy, which is
based on the processing of information, arises naturally within the model.
Our mathematical framework provides a connection between entropy and
fidelity in replication and leads to universal bounds. Error rates similar to
the ones theoretically deduced here in the stepwise equilibrium limit
($\lesssim 10^{-3}$) have been measured for some polymerases, what ultimately
reflects the consistency of this model with the experiments.

Polymerase `pol' and `exo' catalytic residues are conserved throughout
evolution:
viral, prokaryotic and eukaryotic polymerases exhibit common structural
domains and replication mechanisms.
The existence of a certain degree of structural variability and the presence of
replicative complexes, which involve auxiliary proteins and
coordination strategies, should have an effect on fidelity.
In particular, these factors may regulate fidelity to balance
maintenance of genetic identity and the species ability to evolve/adapt,
in most cases increasing fidelity to several order of
magnitude with respect to the values obtained in this work.
Our analysis attempts a universal description of polymerase fidelity since only
basic assumptions common to all polymerases have been made.
This analysis is therefore a starting point for developing
theoretical models describing specific polymerases.
In this regard, highly processive polymerases which do not require a
cooperative association of co-factor proteins like those from some
bacteriophages may constitute the first targets for specific modeling.

Developments of the present analysis for specific polymerases can therefore be
used to test mechanistic hypothesis on polymerase fidelity by contrasting the
subsequently calculated error rates to experimental results. Progresses in this
direction are not only interesting to biology but also to inspire
nanotechnologies in information processing.
Unique to naturally engineered copying/editing nanomachines like the
polymerase enzyme is the inherently stochastic mechanisms by which they manage
classical information, in contrast to artificial devices for which
fluctuations are undesired events.
The theoretical modeling of specific polymerases thus represents a physical
basis to connect classical information processing in biological systems to
artificial, nanoscale platforms, and to open promising avenues in quantum
information copying strategies.


\section*{Appendix A. Hybridization Energies under no neighbor influence\label{sec:G0}}
\noindent
Free energies $\Delta G^{x}_{y}$ are obtained from hybridization energies which include the effect of both
the base-pairing and base-stacking interactions.
The former involve the hydrogen bonding between complementary nucleotides and the latter mainly contain the
hydrophobic interaction between the newly formed base-pair and the previous one.
These energies have been extensively measured and are summarized in~\cite{SantaLucia2004} for both correct, Watson-Crick (WC) base-pairs and mismatches. In the approximation that we are dealing with here, we consider
that these energy levels are degenerate and assume that the hybridization energy only depends on the unmatched nucleotide $y$ in the template strand. Then, we take averages over all previous base-pair possibilities. For this purpose we define next probability distributions for fixed $x$-$y$ base-pairs:

\begin{eqnarray}\label{eq:prob0finecalc}
p_{x, y} (x_0, y_0) & = & \frac{1}{Z_{x, y}}
\exp{\left(\frac{- \Delta G^{x_{o}, x}_{y_{o}, y}}{k T}\right)}, \\
[+3mm]
Z_{x, y} & = & \sum_{x_{o},y_{o} \in {\cal X}}\exp{\left(\frac{- \Delta G^{x_{o}, x}_{y_{o}, y}}{k T}\right)}.
\end{eqnarray}

\noindent
where $\Delta G^{x_{o}, x}_{y_{o}, y}$ is the energy released upon pairing a nucleotide $x$ to another $y$
on the template strand and eventual stacking of the newly formed base-pair with the previous
base-pair made up of a nucleotide $x_0$ in front of another $y_0$. Most of the base-pairs involving two
consecutive non-WC associations are unstable hybridizations and no data were given in~\cite{SantaLucia2004}.
They involve very unfavorable processes (i.e. very high and positive free energies) and we have considered
for these cases that $\Delta G  = +\infty$. Here we use the data at $37^{\circ}C$ and 150 mM NaCl
concentration since polymerization occurs \textit{in vivo} in these conditions.
Then, the hybridization energies in the absence of nearest neighbor interactions are given by

\begin{equation}\label{eq:Ehybrid0}
\Delta G^x_y  = \left\langle \Delta G^{x_{o}, x}_{y_{o}, y} \right\rangle_{x_0,y_0} = 
\sum_{x_{o},y_{o} \in {\cal X}} p_{x, y} (x_0, y_0) \times \Delta G^{x_{o}, x}_{y_{o}, y}.
\end{equation}

Under these considerations, the free energies in $k T$ units are:

\begin{equation}\label{eq:Ehybrid0matrix}
\Delta \textbf{G} = \left(\Delta G^x_y\right) =
\left(\begin{array}{cccc}
 1.01 &  1.70 &  0.34 & -1.65\\
 1.50 &  1.78 & -2.78 &  1.46\\
 0.03 & -2.69 & -0.88 &  0.14\\
-1.64 &  1.39 & -0.16 &  0.71
\end{array}\right),
\\ [+3mm]
\end{equation}

\noindent
where matrix elements follow the order established by the alphabet sequence
$x, y \in {\cal X}  = \{A,C,G,T\}$. Although there is a dependence on both the
temperature and the ionic strength, as explained
in~\cite{SantaLucia2004}, the latter dependence cancels out in the probability
calculation (see Eq.~\textbf{\ref{eq:prob0calc1}}
and~\textbf{\ref{eq:prob0calc2}} in the main text).

\section*{Appendix B. Entropy per nucleotide for the uniform process\label{sec:G0uniform}}

\noindent
The matrix Eq.~\textbf{\ref{eq:prob0matrix}} in the main text can be
interpreted as the probability transition matrix
in a four-state Markov process and thus it is possible to calculate the uniform probability distribution.
By using this distribution in the limit of uniform incorporation of nucleotides when
nearest-neighbor interactions are not taken into account, it is possible to set an upper bound to the
absolute entropy per incorporated nucleotide that is generated in the polymerization of a DNA strand starting from a general template DNA. This calculation is useful because this entropy bound does not depend on the
DNA template sequence. The uniform distribution fulfills the matrix equation $\bf{P} \mbox{\boldmath $\mu$} = \mbox{\boldmath $\mu$}$~\cite{Cover1991} where \mbox{\boldmath $\mu$} is the uniform probability (column) vector. In other words:

\begin{equation}\label{eq:CalcUniform}
\sum_{y \in {\cal X}} p(x||y) \mu (y) = \mu (x)
\end{equation}

\noindent
where $p(x||y)$ and $\mu(x)$ are such that
$\sum_{x \in {\cal X}} p(x||y) = 1$, $\sum_{x \in {\cal X}} \mu (x) = 1$. The entropy per incoporated nucleotide (cf. entropy rate in a random walk) for the uniform process, $s (\cal{X})$, can be
calculated~\cite{Cover1991} from equation

\begin{equation}\label{eq:EntropyUniform}
s ({\cal X}) = -k \sum_{x,y \in {\cal X}} \mu (y) p(x||y) \ln p(x||y).
\end{equation}

By using the probability transition matrix Eq.~\textbf{9} we obtain that the absolute entropy per copied nucleotide in DNA polymerization is bounded by $s ({\cal X}) = 0.643$ $k /nt$ ($37^{\circ}C$). 
This value has been obtained in the limit of, (a), no nearest-neighbor interaction with previously formed base-pairs and, (b), uniform incorporation of nucleotides. This entropic upper bound is independent of the template sequence.

\section*{Appendix C. Partition function vs Markov chain\label{sec:partition}}
\begin{figure}[!ht]
\begin{center}
\includegraphics[draft=false,width=120mm]{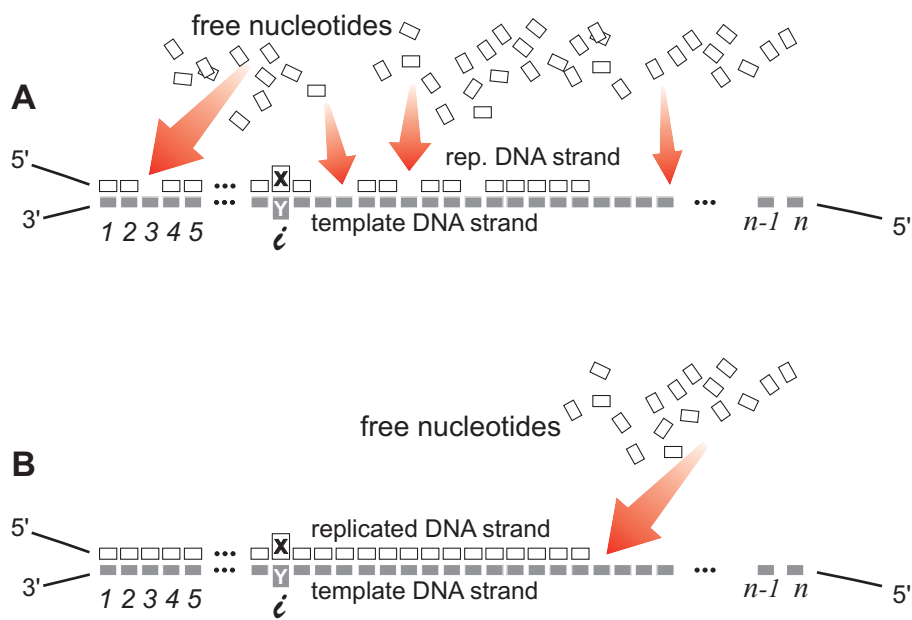}
\end{center}
\caption{{\bf Thermodynamic analysis of the Entropy.}
(\emph{A}) \textit{Ising Mechanism}: Nucleotides
are branched on the template strand without constrainsts of order, direction of replication or number of
nucleotides placed at a time. The calculation is based on the partition function formalism.
(\emph{B}) \textit{Turing Mechanism}: Nucleotides are branched from the 3'-end to the 5'-end of the template
strand on a directional one-after-one basis. The calculation is based on the Markov chain formalism.
\label{fig:SketchMech}}
\end{figure}

\noindent
The state, $\nu$, of the system is specified by a sequence of nucleotides $x_1,\ldots,x_n$ replicated in the
direction 5' to 3' according to a template composed of an ordered sequence of
nucleotides $y_1,\ldots,y_n$ polymerized from the 3'-end to the 5'-end (see
Fig.~\ref{fig:SketchMech}), as denoted by:

\begin{equation}\label{eq:state}
\nu = \left\{ x_1, x_2, \ldots, x_i, \ldots, x_{n-1}, x_n || \mathbf{y} \right\},
\end{equation}


\noindent
where $\mathbf{y} \equiv (y_1,y_2,\ldots,y_i,\ldots,y_{n-1},y_n)$. $x_i$ are variables and $y_i$ parameters
such that $x_i, y_i \in {\cal X}  = \{A,C,G,T\}$. The energy of a state is:

\begin{equation}\label{eq:energy}
E_{\nu} \equiv E \left( x_1,\ldots, x_n || {\bf y} \right) =
\sum_{i=1}^{n} E\left( x_i|x_{i-1}, \ldots, x_1||{\bf y}_{(i)} \right),
\end{equation}

\noindent
where $\mathbf{y}_{(i)} \equiv (y_i|y_{i-1}, \ldots, y_1)$ and $E(x_i|x_{i-1}, \ldots, x_1||\textbf{y}_{(i)})$
is the energy of pairing nucleotide $x_i$ on $y_i$ provided
that nucleotides $(x_1, \ldots x_{i-1})$ are already hybridized on nucleotides $(y_1, \ldots, y_{i-1})$,
respectively. The probability of a state is:

\begin{eqnarray}\label{eq:probability}
P_{\nu} & \equiv & \text{Pr} \{X_1 = x_1, \ldots,  X_n = x_n || {\bf Y} \} = p(x_1, \ldots, x_n|| {\bf y})
\nonumber \\ [+3mm]
& = &
p(x_1||y_1)p(x_2|x_1||y_2|y_1) \cdots p(x_n|x_{n-1}, \ldots, x_1||{\bf y}),
\end{eqnarray}

\noindent
where the last part of the equation is the general expansion of the joint probability as a product of
conditional probabilities~\cite{Fisz1980}.
The mean energy or internal energy of the system is:

\begin{equation}\label{eq:MeanEnergy}
\left\langle E \right\rangle = \sum_{\nu=1}^N P_{\nu} E_{\nu} =
\sum_{x_1, \ldots, x_n} p \left(x_1, \ldots, x_n|| {\bf y} \right)
E \left( x_1, \ldots, x_n || {\bf y} \right),
\end{equation}

\noindent
where $N$ is the number of microstates. The partition function is:

\begin{eqnarray}\label{eq:Z}
Z(\beta, n) & \equiv & \sum_{\nu=1}^{N} \exp{\left( -\beta E_{\nu} \right)} =
\sum_{x_1, \ldots, x_n} \exp{\left( -\beta E\left( x_1, \ldots, x_n||{\bf y} \right) \right)}
\nonumber \\ [+3mm]
& = &
\sum_{x_1, \ldots, x_n} \exp{\left( -\beta \sum_{i=1}^{n}
E\left( x_i|x_{i-1}, \ldots, x_1||{\bf y}_{(i)} \right) \right)},
\end{eqnarray}

\noindent
where $\beta = 1/ k T$. The probability of a configuration is thus:

\begin{eqnarray}\label{eq:ProbIsing}
P_{\nu} & \equiv & Z^{-1}(\beta, n) \exp{\left( -\beta E_{\nu} \right)}
\nonumber \\ [+3mm]
& = &
\frac{ \exp{\left( -\beta \sum_{i=1}^{n} E\left( x_i|x_{i-1}, \ldots, x_1||{\bf y}_{(i)} \right) \right)}}
{\sum_{x'_1, \ldots, x'_n} \exp{\left( -\beta \sum_{j=1}^{n}
E\left( x'_j|x'_{j-1}, \ldots, x'_1||{\bf y}_{(j)} \right) \right)}}
\nonumber \\ [+3mm]
& = &
\frac{ \prod_{i=1}^{n} \exp{ \left( -\beta E\left( x_i|x_{i-1}, \ldots, x_1||{\bf y}_{(i)} \right) \right)}}
{\sum_{x'_1, \ldots, x'_n}
\prod_{j=1}^{n} \exp{ \left( -\beta E\left( x'_j|x'_{j-1}, \ldots, x'_1||{\bf y}_{(j)} \right) \right)}}.
\end{eqnarray}

\noindent
It is important to note that the sums in the denominator over $x'_1,\ldots,x'_n$ are nested and therefore
they cannot be factorized as independent sums. In other words, Eq.~\textbf{\ref{eq:ProbIsing}} expands as:

\begin{eqnarray}\label{eq:ProbIsingExpanded}
& P_{\nu} & = 
\frac{
e^{-\beta E\left( x_1||y_1 \right)}
e^{-\beta E\left( x_2|x_1||y_2|y_1 \right)}
\cdots
e^{-\beta E\left( x_n|x_{n-1}, \ldots, x_1||{\bf y} \right)}
}
{\sum_{x'_1, \ldots, x'_n}
e^{-\beta E\left( x'_1||y_1 \right)}
e^{-\beta E\left( x'_2|x'_1||y_2|y_1 \right)}
\cdots
e^{-\beta E\left( x'_n|x'_{n-1}, \ldots, x'_1||{\bf y} \right)}
}
\nonumber \\ [+3mm]
& = &
\frac{
e^{-\beta E\left( x_1||y_1 \right)}
e^{-\beta E\left( x_2|x_1||y_2|y_1 \right)}
\cdots
e^{-\beta E\left( x_n|x_{n-1}, \ldots, x_1||{\bf y} \right)}
}
{{\sum_{x'_1}} \left[ e^{-\beta E\left( x'_1||y_1 \right)}
 {\sum_{x'_2}} \left[e^{-\beta E\left( x'_2|x'_1||y_2|y_1 \right)}
\cdots
 {\sum_{x'_n}} \left[ e^{-\beta E\left( x'_n|x'_{n-1}, \cdots, x'_1||{\bf y} \right)}
 \right]_{x'_n} \cdots \right]_{x'_2} \right]_{x'_1}
}.
\end{eqnarray}

\noindent
The general term of the expansion of $p(x_1, \ldots, x_n|| {\mathbf y})$ as a product of conditional
probabilities (see Eq.~\textbf{\ref{eq:probability}}) is:

\begin{equation}\label{eq:CondProbZ}
p \left( x_i|x_{i-1}, \ldots, x_1||{\mathbf y}_{(i)} \right) = 
\frac{e^{-\beta E\left( x_i|x_{i-1}, \ldots, x_1||{\mathbf y}_{(i)} \right)}
      f\left( x_1,\ldots,x_{i-1},x_i|| {\mathbf y} \right)}
     {\sum_{x'_i} e^{-\beta E\left( x'_i|x_{i-1}, \ldots, x_1||{\mathbf y}_{(i)} \right)}
      f\left( x_1,\ldots,x_{i-1},x'_i|| {\mathbf y} \right)},
\end{equation}

\noindent
where

\begin{equation}\label{eq:f}
f\left( x_1,\ldots,x_i|| {\mathbf y} \right) \equiv \sum_{x_{i+1},\ldots,x_n}
e^{-\beta E\left( x_{i+1}|x_i, \ldots, x_1||{\mathbf y}^{(i+1)} \right)}
\cdots
e^{-\beta E\left( x_n|x_{n-1}, \ldots, x_1||{\mathbf y} \right)},
\end{equation}

\noindent
fulfilling $\sum_{x_i} p ( x_i|x_{i-1}, \ldots, x_1||{\bf y}_{(i)} ) = 1$. For the last nucleotide in the chain, $i=n$, it follows that $f(x_n||y_n)=1$ and

\begin{equation}\label{eq:CondProbZn}
p \left( x_n|x_{n-1}, \ldots, x_1||{\mathbf y} \right) = 
\frac{e^{-\beta E\left( x_n|x_{n-1}, \ldots, x_1||{\mathbf y} \right)}}
     {\sum_{x'_n} e^{-\beta E\left( x'_n|x_{n-1}, \ldots, x_1||{\mathbf y} \right)}};
\end{equation}

\noindent
but, in general, $f( x_1,\ldots,x_i||y_1,\ldots,y_i ) \neq 1$ and, therefore, the conditional probabilities
depend on the index $i$, i.e. the position in the polymer chain.

The partition function calculation represents an \textit{Ising mechanism} in which the nucleotides are placed arbitrarily: neither order nor one-by-one basis is implied in this replication procedure
(Fig.~\ref{fig:SketchMech}\emph{A}).
Unlike this calculation, the
Markov Chain calculation implies a \textit{Turing Machine mechanism}
in which nucleotides are placed on a one-after-one basis in the 3' to 5' template direction
(Fig.~\ref{fig:SketchMech}\emph{B}).
This mechanism constraints the sequence in which the different configurations are accessible and therefore,
as shown in the main text, the absolute entropy is higher than that from the Ising mechanism.
The probability of placing nucleotide $x_i$ onto $y_i$ within this scheme is given by:

\begin{equation}\label{eq:MarkovProb}
p \left( x_i|x_{i-1}, \ldots, x_1||{\bf y}_{(i)} \right) =
\frac{e^{-\beta E\left( x_i|x_{i-1}, \ldots, x_1||{\bf y}_{(i)} \right) }}
{{\sum_{x'_i}} e^{-\beta E\left( x'_i|x_{i-1}, \ldots, x_1||{\bf y}_{(i)} \right)}}.
\end{equation}

\noindent
Hence, the probability of a configuration $\nu$ is given by:

\begin{eqnarray}\label{eq:ProbMarkovFactorized}
& P_{\nu} & =
p(x_1||y_1)p(x_2|x_1||y_2|y_1) \cdots p(x_n|x_{n-1}, \ldots, x_1||{\bf y})
\nonumber \\ [+3mm]
& = &
\left[ \frac{e^{-\beta E\left( x_1||y_1 \right)}} {{\sum_{x'_1}} e^{-\beta E\left( x'_1||y_1 \right)}} \right]
\times
\left[
      \frac{e^{-\beta E\left( x_2|x_1||y_2|y_1 \right)}}
           {{\sum_{x'_2}} e^{-\beta E\left( x'_2|x_1||y_2|y_1 \right)}}
\right]
\times \cdots \times
\left[
      \frac{e^{-\beta E\left( x_n|x_{n-1}, \cdots, x_1||{\bf y} \right)}}
           {{\sum_{x'_n}} e^{-\beta E\left( x'_n|x_{n-1}, \cdots, x_1||{\bf y} \right)}}
\right]
\end{eqnarray}

\noindent
where we have used brackets to stress that the factorization of the probability here implies that the sums are
independent, unlike in the partitition function calculation, Eq.~\textbf{\ref{eq:ProbIsingExpanded}},
for which the sum could not be factorized. Note that Eq.~\textbf{\ref{eq:ProbMarkovFactorized}} also
fulfills $\sum_{x_1, \ldots, x_n} p(x_1, \ldots, x_n|| {\bf y}) = 1$, which is
a direct consequence of Eq.~\textbf{\ref{eq:MarkovProb}}.

As explained in the main text, hybridization energies are only dependent on the previously formed base-pair,
that is, $l=1$. If boundary changes, such as nucleotide concentration or temperature and ionic gradients along the DNA polymer, are
not present, the energies will only depend on the position, $i$, on the template through the values of $y_i$. This means that there is a single set of energies $\Delta G^{x_{o}, x}_{y_{o}, y}$,
where $x, x_0, y, y_0 \in {\cal X}  = \{A,C,G,T\}$, and parameters $x_0$ and $y_0$ refer to nucleotides
preceding $x$ and $y$, respectively. $\Delta G^{x_o, x}_{y_o, y}$ are the hybridization energy data from~\cite{SantaLucia2004}, also used in previous
\emph{Appendix A}. Note that the fact that the hybridization energies do not
depend on $i$ implies that there is also single set of probabilities for the
Turing mechanism:

\begin{eqnarray}\label{eq:prob1finecalc}
p(x|x_0||y|y_0) & = & \frac{1}{Z(x_0,y,y_0)}\exp{\left(\frac{- \Delta G^{x_o, x}_{y_o, y}}{k T}\right)}, \\
[+3mm]
Z(x_0,y,y_0)  & = & \sum_{x \in {\cal X}}\exp{\left(\frac{- \Delta G^{x_{o}, x}_{y_{o}, y}}{k T}\right)}.
\end{eqnarray}

\noindent
The dependence on the salt cancentration cancels out in this calculation, as
was shown for the $0^{th}$ order approximation (see
Eqs.~\textbf{\ref{eq:prob0calc1}} and~\textbf{\ref{eq:prob0calc2}}), so these
probability distributions
are only temperature-dependent within the empirical expressions given by~\cite{SantaLucia2004}.

For the Ising mechanism, however, the probabilities are affected not only by
the sequence parameters $y_i$ but also by the length, $n$, of the template,
and therefore, there is an implicit dependence on $i$ (compare
Eqs.~\textbf{\ref{eq:CondProbZ1}} and~\textbf{\ref{eq:MarkovProb1}} in the
main text or
Eqs.~\textbf{\ref{eq:CondProbZ}} and~\textbf{\ref{eq:MarkovProb}}).

\section*{Appendix D. Montecarlo simulations\label{sec:MontecarloSim}}
\noindent
Simulations of the Internal energy and the probability of error
for both the Ising and Turing mechanisms have been generated for
template sequences of 20 nucleotides.
For the former, the partition function is calculated by using the classical Metropolis-Montecarlo
procedure~\cite{Chandler1987}.
For the Turing mechanism, each decision step is ruled by the probabilities
$p \left( x_i|x_{i-1}||y_i|y_{i-1} \right)$, as established by
Eq.~\textbf{\ref{eq:MarkovProb1}} in the main text.
More in depth, a symbol $x_i \in {\cal X}  = \{A,C,G,T\}$ and a number $r \in [0,1]$ are chosen at random at
step $i$. Then, considering the outcome of previous
step $i-1$ and the template symbol $y_i$, the symbol $x$ incorporated at position $i$ is that fulfilling
the condition $r < p \left( x_i|x_{i-1}||y_i|y_{i-1} \right)$.

For both kind of calculations, averages $\langle E \rangle$ are taken over $10^7$ iterations, i.e. over $10^7$ sequences generated with each procedure. Comparison of the exact values shown in
Fig.~3\emph{A} with the ones
simulated here show a discrepancy within 3.6\%, thus confirming a rapid convergence to the
thermodynamic limit. The calculations for independent, non-identically distributed variables show discrepancies
from the exact calculations in the fifth significant figure for the simulations based on the above procedure,
which confirms the validity of the above explained algorithm.

\section*{Appendix E. \textit{Informational} and \textit{Configurational} terms
of the Entropy in an ideal gas
\label{sec:IdealGasEntropy}}
\noindent
The entropy per particle of a multicomponent ideal gas can be calculated by using the Sackur-Tetrode formula
for a mixture of $n$ particles of types $x \in {\cal X} = \{A,C,G,T\}$ and masses $m_x$ occupying a total
volume $V$ as follows:

\begin{eqnarray}\label{eq:Sackur}
s & \equiv & \frac{S}{n} = \sum_{x \in {\cal X}} \left\{ \frac{n_x}{n} k \ln \frac{V}{n_x} + \frac{3}{2}
\frac{n_x}{n} k
\left[ \frac{5}{3} +\ln \left( \frac{2 \pi m_x k T}{h^2} \right) \right]    \right\}
\nonumber \\ [+3mm]
& = &
- k \sum_{x \in {\cal X}} p_n(x) \ln p_n(x) - k \sum_{x \in {\cal X}} p_n(x)
\ln \left( \frac{n h^3}{V \left( 2 \pi m_x k T \right) ^{3/2}} \right) + \frac{5}{2} k
\nonumber \\ [+3mm]
& = &
s_I(n_x/n) + s_C(x,C) + s_0,
\end{eqnarray}

\noindent
where $p_n(x) = n_x / n$, as defined in the main text. $s_0= (5/2) k$ is a
constant, $s_I(n_x/n)= -k \sum_{x \in {\cal X}} p_n(x)\times$ $\ln p_n(x)$ is the herein labeled as \textit{informational entropy} of the initial state
and $s_C(x,C)$ is the herein labeled as \textit{configurational entropy},

\begin{equation}\label{eq:EntropyConf}
s_C (x,C) = - k \sum_{x \in {\cal X}} p_n(x)
\ln \left( C \Lambda _l ^3 (x)\right),
\end{equation}

\noindent
which depends on the type of particle, $x$, and the total concentration of particles, $C=n/V$.
$\Lambda_l(x)$ is a \textit{generalized thermal wavelength} for particles in a liquid, introduced here as:

\begin{equation}\label{eq:GenThermWave}
\Lambda_l(x) \equiv \frac{1}{n_l^T (x)} \frac{h}{\sqrt{2 \pi m_x k T}},
\end{equation}

\noindent
with $n_l^T (x)$ a \textsl{thermal refractive index} of particles, $x$, in a liquid, which becomes
$n_{{vac}}^T = 1$ for particles in vacuum, since this is the case of the classical ideal gas.
In this scheme, the thermal refractive index modifies the thermal wavelength of classical particles in a liquid as $\Lambda_l = \Lambda/ n_l^T$, and the dispersion relation as $E = (n_l^T)^{-2} p^2/2m$.

For particles of similar mass, chemical composition and structure, as it is the case of nucleotides,
we can assume that both $\Lambda_l$ and $n_l^T$ are the same for the four types of particles. In these
conditions, the configurational entropy is only a function of the total concentration of nucleotides,
$s_C (C) = - k \ln \left( C \Lambda _l ^3 \right)$.
The molar concentration of nucleotides in a reservoir for single-molecule experiments and for
\textit{in vivo} replication is certainly of $\sim 50$ $\mu M$. The mass of the deoxyribonucleotide
monophosphates is $m_x \simeq 330$ $Da$.
Then, the thermal wavelength is $\Lambda \sim 10^{-12}$ $m$ and the
configurational entropy is $s_C \simeq 26$ $k/nt$, which is much larger than the informational
entropy, $s_I = 1.39$ $k/nt$ (see the main text).

\section*{Acknowledgments}
It is a pleasure to acknowledge that numerous stimulating
discussions ---including those with B. Ibarra, F. Ritort, J.M. Parrondo,
E. Roldan and M. Ribezzi--- contributed to the derivation and understanding
of the results presented in this paper. The author also wishes to thank
J.M. Valpuesta and J.L. Carrascosa for support and encouragement in
this research.\\

\newpage

\bibliography{entropy}{}

\section*{Figure Legends}

\end{document}